\documentclass[12pt]{article} 

\usepackage{amsfonts}
 \usepackage{amssymb}
 \usepackage{amsmath}

\def\pmx{\begin{pmatrix}}
\def\emx{\end{pmatrix}}
\def\bsq{\begin{subequations}}
\def\esq{\end{subequations}}

\newtheorem{thm}{Theorem}

\newtheorem{qes}[thm]{Question}

\def\be{\begin{eqnarray}}
\def\ee{\end{eqnarray}}
\def\bee{\begin{eqnarray*}}
\def\eee{\end{eqnarray*}}

\def\bra{\langle}
\def\ket{\rangle}
\def\kb{ \ket \bra }

\newcommand{\kt}[1]{ | #1 \ket }
\newcommand{\proj}[2]{ | #1 \kb #2 | }

\def\dg{\dagger}

\def\holv{{\rm  Holv}}
\def\eof{{\rm EoF}}

\def\opt{{\rm opt}}

\def\half{{\textstyle \frac{1}{2}}}
\def\1rt2{{\textstyle \frac{1}{\sqrt{2}}}}

 \def\tr{{\rm Tr}}

\def\ot{\otimes}

\def\wtd{\widetilde}

\def\cale{{\cal E}}
\def\calb{{\cal B}}
\def\hil{{\cal H}}

\def\nn{\nonumber}

          \title{Some Bipartite States Do Not Arise from Channels}

        \author{Mary Beth Ruskai\thanks{Work partially supported  by
 the U.S. National Science  Foundation under Grant number DMS-03-14228.}
\\ Department of Mathematics, Tufts University\\
  Medford, Massachusetts 02155 USA \\ 
 {\normalsize marybeth.ruskai@tufts.edu}}

 \date{\today \\ ~~ \\
 Dedicated to Charles H. Bennett on his 60th Birthday }

\begin{document}

\maketitle

\begin{abstract}
   It is well-known that the action of a quantum channel on a state can be
represented, using an auxiliary space,  as the partial trace of an
associated bipartite state.   Recently, it was observed that
for the bipartite state associated with the optimal average
input of the channel, the entanglement of formation is simply
the entropy of the reduced density matrix minus the Holevo capacity.
It is natural to ask if every bipartite state can be associated
with some channel in this way.  We show that the answer is
negative.
\end{abstract}

\noindent{\small PACS number 03.67;  MSC classification 82P68.}


\section{Background}

Recently, Matsumoto, Shimono and Winter (MSW) \cite{MSW} pointed out
an important connection between the  channel
capacity and of entanglement of formation, which allows one
to draw some conclusion about the additivity of the latter
from that of the former.   There has
also been speculation that the  additivity of   channel
capacity and of entanglement of formation, are equivalent.
A connection between   additivity of entanglement of formation
and multiplicativity of the p-norm measure of purity has
also been given by Audenaert and  Braunstein \cite{AB}.

In view of this it is natural to ask if every bipartite state
 $\rho_{AB}$ can be associated with a channel in the sense of MSW.
In particular is its reduced
density matrix, $\rho_{A}$,  the optimal average output of a channel 
whose capacity is related  to the entanglement of formation of $\rho_{AB}$ 
as described in \cite{MSW}?   We  will give a more precise formulation of
this statement and show that the answer is negative.

Recall that a state is represented by a density matrix $\rho$,
i.e., a positive semi-definite operator with $\tr \rho = 1$.
The von Neumann entropy of $\rho$ is $S(\rho) = -\tr \rho\log \rho$.
By an ensemble $\cale = \{ \pi_i, \rho_i \}$ we means a set of density
matrices, $\rho_i$ and associated probabilities $\pi_i$.
By a channel $\Phi$ we mean a completely positive, trace-preserving
(CPT) map.  
The {\em Holevo capacity} of the map $\Phi$ is
\be  \label{eq:holv}
   C_{\holv}(\Phi) = \sup_{\cale} \, \Big\{ S[ \Phi(\rho)] -
     \sum_i \pi_i S[ \Phi(\rho_i)] \Big\}
\ee
where the supremum is taken over all ensembles 
$\cale = \{ \pi_i, \rho_i \}$ and $\rho = \sum_i \pi_i \rho_i$.
The {\em optimal average input} is the state
$\rho_{\opt} = \sum_i \pi_i \rho_i$ associated with the ensemble which
attains the supremum in (\ref{eq:holv}).  The 
 optimal average {\em output} is then $\Phi(\rho_{\opt})$.

The key to the MSW construction is the following result
of Stinespring \cite{S}, which was subsequently used by
Lindblad  \cite{L} in his work on relative entropy.
\begin{thm}   \label{thm:stine} 
Given a CPT map $\Phi$ on $\calb(\hil)$ , one can find an auxiliary
 space $\hil_B$, a density matrix $\nu_B$ and a unitary map
$U_{AB}$ on $\hil_A \ot \hil_B$ such that
\be  \label{eq:stine}
    \Phi(\rho) = \tr_B \big[ U^{\dg}_{AB} \rho \ot \nu_B U_{AB} \big]
\ee
where $\tr_B$ denotes the partial trace and we have identified
$\hil$ with $ \hil_A$.
\end{thm}

Although we are interested in the case in which $\rho$ is a 
density matrix, the representation (\ref{eq:stine}) is valid 
for all operators in $\calb(\hil)$.  When
$\Phi(\rho) = \sum_k A_k^{\dg} \rho A_k$, it is easy to construct a
representation of the form (\ref{eq:stine}), as
discussed in section III.D of \cite{R}.   In   fact,
$U^{\dg}_{AB} \rho \ot \nu_B U_{AB}$ is the block matrix
which has the form $\sum_{jk} A_j^{\dg} \rho A_k \ot \proj{j}{k}$.

Given a bipartite state $\gamma_{AB}$ on $\hil_A \ot \hil_B$, its
entanglement of formation satisfies
\be
   \eof(\gamma_{AB}) = \inf \Big\{ \sum_k \pi_k S\big(\tr_B \gamma_k
  \big) \, : \, \{ \pi_k \gamma_k \} \hbox{ ensemble with } 
   \gamma_{AB} = \sum_k \pi_k \gamma_k \Big\}.
\ee
Although it is customary to define the \eof  ~using ensembles
for which all $\gamma_k$ are pure states, there is no loss
of generality in allowing arbitrary states.

\begin{thm} {\rm (Matsumoto, Shimono and Winter)}
Let $\Phi$ be a CPT map with a representation $\hil_B, \nu_B, U_{AB}$
as in Theorem~\ref{thm:stine}.  
The Holevo capacity of $\Phi$ satisfies
\be  \label{eq:MSW}
  C_{\holv}(\Phi) = \sup \Big\{ S\big( \tr_B \gamma_{AB} \big) -
    \eof(\gamma_{AB}) \, : \,  
  \gamma_{AB} = U^{\dg}_{AB} \rho \ot \nu_B U_{AB} \Big\}
\ee
where $\rho$ is a density matrix on $\hil = \hil_A$.   Moreover,
the state $\wtd{\gamma}_{AB}$ which attains this supremum 
satisfies $\tr_B \, \wtd{\gamma}_{AB} = \Phi(\rho_{\opt})$ where
$\rho_{\opt}$ is the optimal input of $\Phi$.
\end{thm}
It follows immediately that for the state $\wtd{\gamma}_{AB}$
\be   \label{eq:conj}
  \eof(\wtd{\gamma}_{AB}) =    
   S\big[\Phi( \rho_{\opt})\big] - C_{\holv}(\Phi).
\ee  
It is thus natural to ask the following
\begin{qes} \label{Q}
 Given a bipartite state $\gamma_{AB}$ on $\hil_A \ot \hil_B$, is there
a CPT map $\Phi$ on $\calb(\hil_A)$ such that $\tr_B \gamma_{AB}$
is the optimal average output state of $\Phi$ and
\be
  \eof(\gamma_{AB}) = S\big( \tr_B \gamma_{AB} \big) - C_{\holv}(\Phi)? 
\ee
\end{qes}
We will show that the answer to this question is negative.

Our counter-example is based on a result
in \cite{OPW} and \cite{SW} 
which comes from the fact that when $\rho = \sum_k \pi_k \rho_k$
\be
   S(\rho) - \sum_k \pi_k S(\rho_k) = \sum_k \pi_k H(\rho_k,\rho)
\ee
where the {\em relative entropy} is defined as
\be  \label{thm:OPW}
  H(\omega,\rho) = \tr \, \omega \big[ \log \omega - \log \rho \big].
\ee
\begin{thm}  \label{thm}
  Let $\Phi$ be a CPT map with optimal average input $\rho_{\opt}$.  Then
\be
  C_{\holv}(\Phi) = \sup_{\omega} H[ \Phi(\omega), \Phi(\rho_{\opt})] .
\ee
\end{thm}      
It follows as an immediate corollary, that if 
$\cale = \{ \pi_k, \rho_k \}$ is any
optimal ensemble for the channel $\Phi$, then
$H[ \Phi(\rho_k), \Phi(\rho_{\opt})] = C_{\holv}(\Phi)$ is independent of
$k$, i.e., the optimal average output is ``equi-distant'' in the 
sense of relative entropy from all outputs $\Phi(\rho_k)$ in the
 ensemble.

\section{Counter-example}

An affirmative answer to Question~\ref{Q} above would 
imply that if the ensemble $\{ \pi_k, \proj{\psi_k}{\psi_k} \}$ is optimal
for $\eof(\gamma_{AB})$, then 
$\{ \pi_i, \tr_B \proj{\psi_k}{\psi_k} \}$ would
be an optimal {\em output} ensemble for the corresponding map 
$\Phi$.  It would then follow from  the equi-distance corollary to
Theorem~\ref{thm:OPW} that
\be  \label{eq:cont}
 H( \omega_k, \gamma_A) = 
  -S(\omega_k ) -  \tr_A \, \omega_k \log \gamma_A = C.
\ee
where $\omega_k = \tr_B \proj{\psi_k}{\psi_k}$ and $C$ is a constant which
is independent of $k$.
(In fact, $C$ is the Holevo capacity of $\Phi$ if such a channel exists.)
We will present an example of a bipartite qubit state which
does not satisfy (\ref{eq:cont}).

It may be that counter-examples to (\ref{eq:cont}) can already
be found in the literature on entanglement.   However, we will
take advantage of a result of Wootters \cite{W} to construct a
rather simple qubit counter-example which does not require 
extensive numerical computation.   In \cite{W}, Wootters obtained
an explicit formula for the $\eof$ of a bipartite qubit
state using a quantity called the concurrence.   Moreover, he
showed that the $\eof$ could be achieved using an ensemble of
at most four pure states $\proj{\psi_k}{\psi_k}$, for which 
$S(\tr_B \proj{\psi_k}{\psi_k}) = \eof(\gamma_{AB})$.  Thus, for such an
ensemble, (\ref{eq:cont}) holds if and only if 
\be  \label{eq:A}
   \tr_A [(\tr_B \proj{\psi_k}{\psi_k}) \log \gamma_A]
\ee
is independent of $k$.  We can assume  without loss of generality  that
$\gamma_A$ is diagonal with eigenvalues $\half(1 \pm x)$.  Then
$\tr \, \omega \log \gamma_A$ depends only on the diagonal
elements of $\omega$ which can be written as 
$\half(1 \pm d)$.  In fact,
\be
   \tr \omega \log \gamma_A  & = &
   \tfrac{1+d}{2} \log \tfrac{1+x}{2} +
   \tfrac{1-d}{2} \log \tfrac{1-x}{2} \\ \nn
 & = & \log 2 - \half \,
   \big[ \log(1-x^2) + d \log \tfrac{1+x}{1-x} \, \big]  
\ee
which depends linearly on $d$ for fixed $x \neq 0$.   Thus, 
when $\gamma_A \neq \half I$, (\ref{eq:A}) is independent
of $k$ if and only if all $\omega_k = \tr_B \proj{\psi_k}{\psi_k}$ have
the same diagonal elements.   However, we also know that
$S(\omega_k) = \eof(\gamma_{AB})$ is independent of $k$, which
implies that all $\omega_k$ have the same eigenvalues.
Thus, all the reduced density matrices $\tr_B \proj{\psi_k}{\psi_k}$ must
have the form $\half \pmx 1 +d & e^{i \theta_k} t \\
     e^{-i \theta_k} t & 1-d \emx $ for some fixed $t$.
It is not hard to find an example for which this does not
hold.

Let $\kt{\beta_0} = \1rt2 ( \kt{00} + \kt{11} )$ and 
$\kt{\beta_3} = (\sigma_z \ot I)\kt{\beta_0} = 
  \1rt2 ( \kt{00} - \kt{11})$ be the indicated
maximally entangled Bell states, and
\be  \label{eq:cntex}
  \gamma_{AB} = \tfrac{5}{8} \proj{\beta_0}{\beta_0} + 
     \tfrac{1}{16} \proj{\beta_3}{\beta_3} + \tfrac{1}{4} \proj{01}{01}
    + \tfrac{1}{16} \proj{10}{10} .
\ee
The reduced density matrix is 
  $\gamma_A = \tr_B \gamma = \half(I + \frac{3}{16} \sigma_z) \neq \half
I$. One can show that the optimal $\eof$ decomposition of (\ref{eq:cntex})
has the form $\gamma_{AB} = \sum_{k = 1}^4 \pi_k \proj{\psi_k}{\psi_k}$
with $\kt{ \psi_1} = a_0 \kt{\beta_0} + a_3 \kt{\beta_3}$ so that
$\omega_1 = \tr_B \, (\proj{\psi_1}{\psi_1}) $ is diagonal
 in the basis $\kt{0}, \kt{1}$
(i.e, $t = 0$ in the matrix above).
However, the remaining $\psi_k$ are superpositions which contain
 Bell states $\kt{\beta_k}$ and the product states
$ \kt{01}, \kt{10}$ in a form which necessarily yields a 
reduced density matrix $\omega_k$ which is {\em not} diagonal.  By the
discussion above, this implies a negative answer to Question~\ref{Q}.

To actually find the entanglement of formation and optimal
decomposition of $\gamma_{AB}$, let
\be
 \wtd{\gamma} & = & \tfrac{5}{8} \proj{\beta_0}{\beta_0} + 
     \tfrac{1}{16} \proj{\beta_3}{\beta_3} + \tfrac{1}{4} \proj{10}{10}
    + \tfrac{1}{16} \proj{01}{01}
\ee
be the density matrix with all spins flipped.
The concurrence $\mu$ can be expressed in terms of the
eigenvalues of $( \sqrt{\gamma} \, \wtd{\gamma} \, \sqrt{\gamma} )^{1/2}$.
Following Wootters \cite{W}, one finds $
\mu = \frac{5}{8} - \frac{1}{16} - \frac{1}{8} - \frac{1}{8} =
\frac{5}{16}$.  Let
\be
 h(x) =  - \tfrac{1+x}{2} \log  \tfrac{1+x}{2}
    - \tfrac{1-x}{2} \log  \tfrac{1-x}{2}.
\ee
Then, proceeding  as described in \cite{W}, one finds
\bee
  \eof(\gamma_{AB}) = h(\sqrt{1 - \mu^2}) =
      h \Big(\tfrac{\sqrt{231}}{16} \Big) \approx 0.1689 ~ <  ~ 
  0.9745 \approx h\Big(\tfrac{3}{16}\Big) = S(\gamma_A ).
\eee 
The optimal ensemble has weights
\bee
  \pi_1 = 0.1527 ,~ \pi_2 = \pi_3 = \pi_4 =   0.2824
\eee
associated with the projections for the following pure states
\bee
  \kt{ \psi_1} & = &  ~~0.8101 \kt{\beta_0} +  0.5863 \kt{\beta_3}  \\ 
 \kt{ \psi_2} & = &  
-0.7870 \kt{\beta_0} +  0.1087 \kt{\beta_3} + 0.5432 \kt{01} + 0.2716
\kt{10}
\\ 
 \kt{ \psi_3} & = & ~~0.7870  \kt{\beta_0}  - 0.1087 \kt{\beta_3} +
 0.5432  e^{i \pi/3} \kt{01} +   0.2716e^{-i \pi/3} \kt{10}  
\\
 \kt{ \psi_4} & = &  -0.7870  \kt{\beta_0}  + 0.1087 \kt{\beta_3} 
- 0.5432  e^{-i \pi/3} \kt{01} -    0.2716 e^{i \pi/3} \kt{10} 
\eee

One can easily verify that the diagonal elements of 
$\omega_1 = \tr_B \, \proj{\psi_1}{\psi_1} $ are not equal
to those of $\omega_k = \tr_B \, \proj{\psi_k}{\psi_k} $
for $k = 2, 3, 4$.  One can also compute the reduced
density matrices  $\omega_k$ and see that
  $\omega_1$ is diagonal in the basis $\kt{0}, \kt{1}$,
but the others are not.   Alternatively,
one can observe that if a reduced density matrix is diagonal in the basis
$\kt{0}, \kt{1}$, then any purification must have the form
\be  \label{eq:schmidt}
  \kt{\Psi} = a \kt{0} \ot \kt{\phi_1} + b \kt{1} \ot \kt{\phi_2}.
\ee
with $\kt{\phi_1}, \kt{\phi_2}$ orthogonal.  
By rewriting
\bee
  \kt{ \psi_2} & =  &  
- 0.4796 \, \kt{00}  -0.6333\,  \kt{11} + .5432 \, \kt{01}  + .2716 
   \, \kt{10}
\\
 & =  &   - \kt{0} \ot \big( 0.4796 \, \kt{0} - .5432 \,  \kt{1} \big)  
  + \kt{1} \ot  \big( .2716  \, \kt{0} -   0.6333 \, \kt{1} \big) ,
\eee
one sees that $\psi_2$ does not have the form (\ref{eq:schmidt})
with orthogonal $ \kt{\phi_k}$.  The actual reduced density matrices 
have the form
$\omega_1 = \half \big[ I + \frac{\sqrt{231}}{16}   \sigma_z \big] =
   \pmx 0.9750 & 0 \\ 0 & 0.0250 \emx$,
  $\omega_k = \pmx 0.5251 & e^{i \theta_k }0.4743 \\ 
   e^{-i \theta_k} 0.4743 & 0.4749
\emx $ with $\theta_2 = \pi, \theta_3 = \frac{\pi}{3}, 
  \theta_4 = - \frac{\pi}{3}$.

\section{Remarks on representations}

In the canonical method of constructing a representation of 
the form (\ref{eq:stine}), the reference state $\nu_B$ is pure
and $d_B$ the dimension of $\hil_B$ is equal to the number of
Kraus operators.  Then the lifted state
$\gamma_{AB} = U_{AB} \rho \ot \nu_B U_{AB}^{\dg}$ for which
$\Phi(\rho) = \tr_B \gamma_{AB}$ has rank at most $d = d_A$, 
the dimension of the original Hilbert space.   This is clearly
a very restricted class of bipartite states.    Moreover,
generically, $d_B > d$ since many maps require the maximum
number $d^2$ of Kraus operators.   Thus the canonical 
representation yields only bipartite states which are far
more singular than typical states.

Some maps $\Phi(\rho)$ can be represented in the form 
(\ref{eq:stine}) using a mixed reference state $\nu_B$.
Indeed, given a mixed bipartite state $\nu_B$ and unitary
operator $U_{AB}$, (\ref{eq:stine}) can be used to define
a channel $\Phi$.   Unfortunately, relatively little is known
about such mixed state representations.
It has been suggested that one might be able to represent
a channel $\Phi$ in the form (\ref{eq:stine}) using a space
of dimension $d_B = d$ if mixed states are used.   However,
this is known not to be true in general \cite{T,ZR}.

Now suppose that   a qubit channel $\Phi$ can be represented
(possibly using a mixed reference state $\nu_B$) in the form
(\ref{eq:stine}) using an auxiliary qubit space $\hil_B$ 
so that $d_B = 2$.  Then the argument given before (\ref{eq:A})
can be used to show that for the optimal input distribution
$\{ \pi_k \rho_k \}$  
\be   \label{eq:B}
   \tr \Phi(\rho_k) \log \Phi(\rho_{\opt})
\ee
must be independent of $k$.  One can easily check that the
3-state examples given in \cite{KNR} do {\em not} have
this property.  Most     non-unital qubit maps for which the
translation of the image of the Bloch sphere lacks symmetry
will also violate (\ref{eq:B}).   This implies that such
channels  require an auxiliary space with $d_B > 2$.
One expects $d_B = 4$, consistent with the fact that such
maps also require 4 Kraus operators.
This gives another, somewhat indirect, proof that some
CPT maps require an auxiliary space with $d_B > d$.

It would be of some interest to characterize the maps
which admit mixed state representations with $d_B = d$,
as well as the bipartite states corresponding to the
optimal average output.

\medskip

\noindent{\bf Acknowledgment:}  
It is a pleasure to thank Professor Christopher King
for stimulating and helpful discussions.

 \pagebreak


\end{document}